\documentclass{vgtc}                          

\graphicspath{{figures/}{pictures/}{images/}{./}} 

\usepackage{times}                     

\usepackage{tabu}                      
\usepackage{booktabs}                  
\usepackage{lipsum}                    
\usepackage{mwe}                       
\usepackage{amsmath}
\usepackage{mathptmx}                  
\usepackage{kotex}
\usepackage{listings}
\usepackage{fancyvrb}

\usepackage{tikz}
\newcommand{\circled}[1]{%
  \tikz[baseline=(char.base)]{%
    \node[shape=circle, fill=black, text=white, inner sep=1pt, font=\footnotesize\bfseries] (char) {#1};%
  }%
}

\renewcommand{\paragraph}[1]{\vspace{4pt}\noindent\textbf{#1.}}

\newsavebox{\FVerbatimbox}
\newlength{\FVerbatimwidth}
\AtBeginDocument{%
  \sbox0{\ttfamily X}%
  \setlength{\FVerbatimwidth}{60\wd0}%
  \addtolength{\FVerbatimwidth}{2\fboxsep}
}

\onlineid{1009}

\vgtccategory{Research}

\vgtcinsertpkg

\title{A Multi-Level Visual Analytics Approach to Artist–Era \\ Alignment in Popular Music}

\author{
Jiyeon Bae \thanks{e-mail: jybae@hcil.snu.ac.kr}  \quad
Jinwook Seo\thanks{e-mail: jseo@snu.ac.kr}  \\ 
Seoul National University \\
}

\teaser{
  \vspace{-0.2cm}
  \centering
  \includegraphics[width=\linewidth]
  {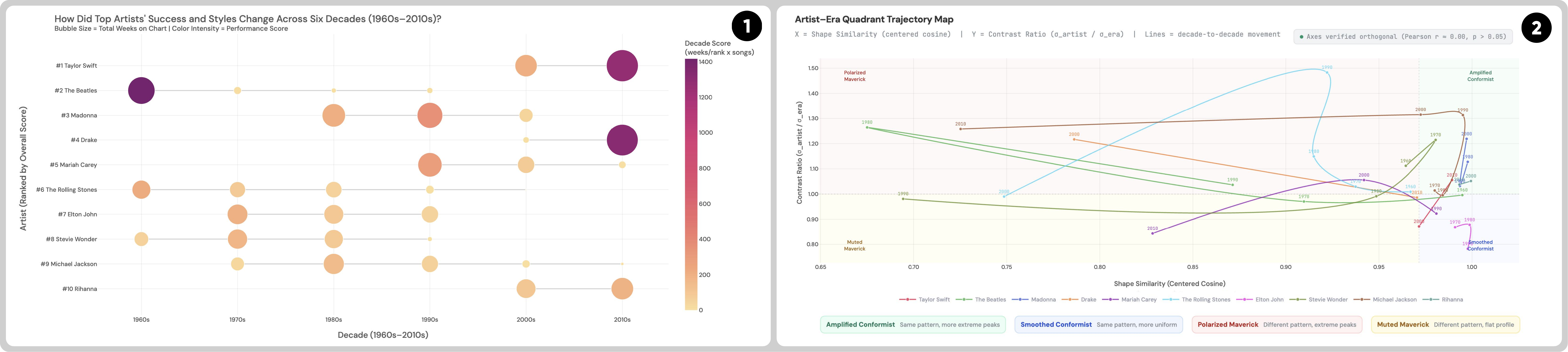}
  \vspace{-0.6cm}
  \caption{System overview. (1) Artist--Decade Bubble Chart 
  showing performance trajectories across six decades. 
  (2) Quadrant Trajectory Map plotting artist--decade units 
  along shape similarity and contrast ratio. Detail views 
  are shown in Figure~\ref{fig:detail}.}
  \label{fig:main}
}

\abstract{
Existing computational studies of popular music primarily 
model aggregate trends or predict chart performance, 
offering limited support for interpreting artist-level 
alignment against historical stylistic baselines. We 
introduce an interactive visual analytics framework that 
treats each artist--decade as a unit defined relative to 
an era-specific baseline, characterized along two 
complementary dimensions: profile shape similarity, 
capturing directional correspondence with the era's 
feature pattern, and profile contrast ratio, capturing 
stylistic intensity relative to the era's dispersion. 
Together, these dimensions define a quadrant-based 
trajectory space for reasoning about conformity, divergence, and stylistic amplification or attenuation. Applied to weekly 
U.S. \textit{Billboard Hot 100} chart entries from the all-time 
top-10 artists across six decades (1960s--2010s), linked 
with Spotify audio features, the framework reveals that 
alignment and intensity can meaningfully diverge across 
artist trajectories.
}


\begin{document}
\maketitle

\section{Introduction}

Understanding how artists position themselves relative to evolving 
mainstream norms is central to analyzing longitudinal stylistic 
change in popular music. The \textit{Billboard Hot 100} provides a 
historically grounded record for examining such change.

Prior computational studies have largely followed two directions: 
popularity prediction and aggregate temporal analysis. 
Prediction-oriented work has examined associations between audio 
features and chart performance~\cite{Soares19, Li24}, though these 
models typically treat datasets as temporally homogeneous. Mauch et 
al.~\cite{Mauch2015} detected discrete stylistic revolutions in U.S. popular music through topic-based audio analysis. In visual 
analytics, prior systems have supported music exploration through 
embedding- and timeline-based views~\cite{Chang23, Levesque21}. 
While effective, these approaches operate at aggregate or 
discrete-cluster levels and frame analysis around proximity-based 
relationships, leaving the continuous positioning of individual 
artists relative to era-specific norms underexplored.

Existing approaches do not separate the \emph{direction} 
of stylistic alignment from its \emph{intensity}. To address this 
gap, we introduce an interactive visual analytics framework that 
operationalizes each \emph{artist--decade} relative to an era 
baseline. Each profile is characterized along two complementary 
dimensions: \emph{shape similarity}, capturing directional 
correspondence with the era's feature pattern, and \emph{contrast 
ratio}, capturing stylistic intensity relative to the era's 
dispersion. Together, these dimensions define a quadrant-based 
trajectory space for analyzing stylistic trajectories over time. 
Applied to weekly U.S. \textit{Billboard Hot 100} chart entries 
from the all-time top-10 artists across six decades 
(1960s--2010s), linked with Spotify audio features, the framework 
enables exploration from era baselines to artist trajectories and 
song-level deviations, revealing alignment configurations not 
discernible through aggregate analysis alone.

\section{Interactive Visualization}
\vspace{-0.1cm}
\paragraph{Analytic Tasks}
Informed by a formative interview with an expert, the system 
supports three analytic tasks from overview to detail:
\vspace{-0.16cm}
\begin{itemize}
  \item \textbf{T1:} Trace artist trajectories across decades.
  \vspace{-0.26cm}
  \item \textbf{T2:} Examine song-level performance and audio profiles.
  \vspace{-0.26cm}
  \item \textbf{T3:} Quantify and compare artist--era stylistic alignment.
\end{itemize}
\vspace{-0.26cm}

\begin{figure}[tb]
  \centering
  \includegraphics[width=\linewidth]{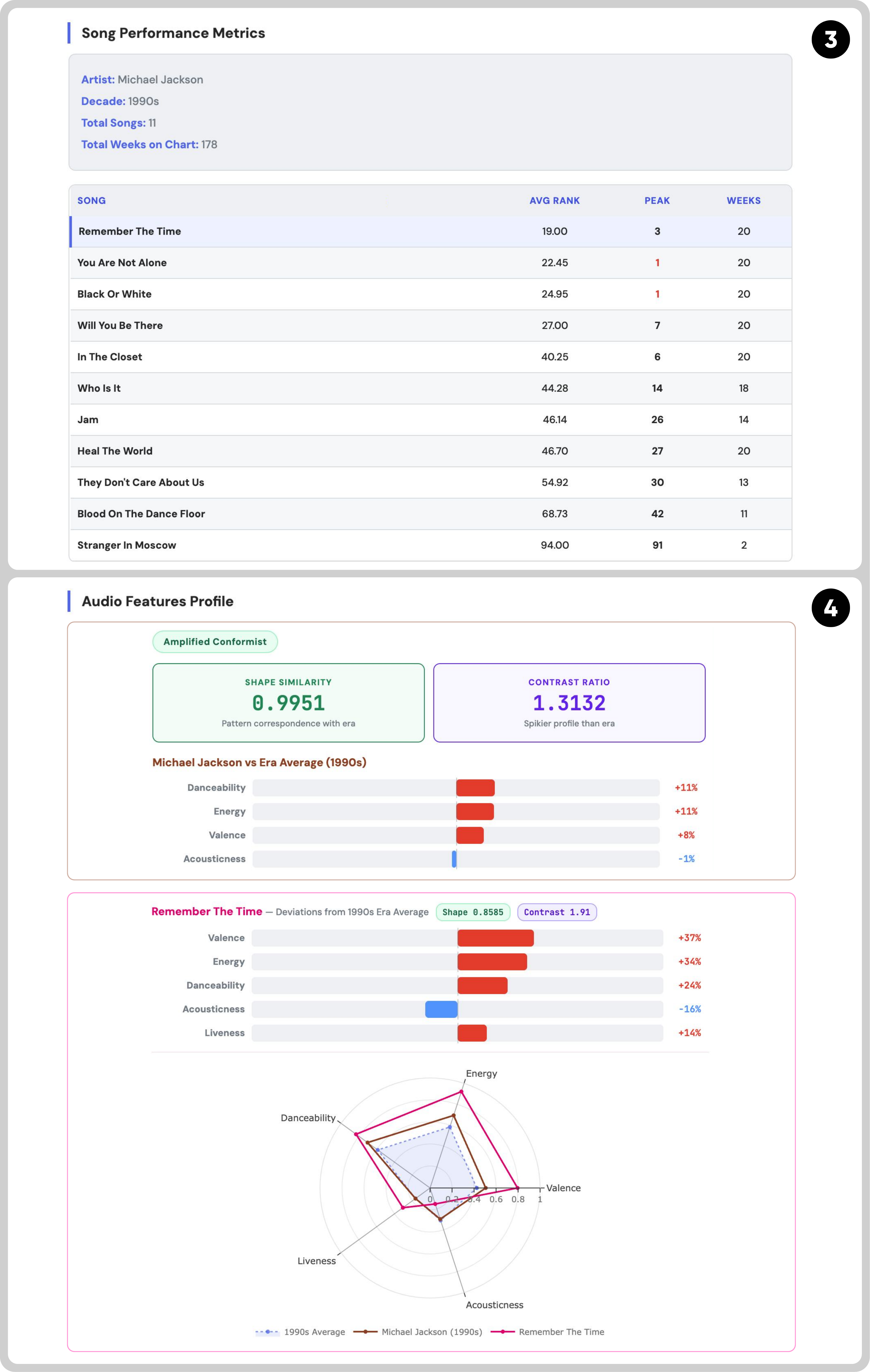}
  \vspace{-0.6cm}
  \caption{Michael Jackson (1990s) and ``Remember The Time'' 
  selected. (3) Song Performance Table listing charting metrics. 
  (4) Audio Features Profile displaying quadrant classification, 
  per-feature deviations from the era average, and a song-level 
  radar overlay.}
  \label{fig:detail}
    \vspace{-0.5cm}
\end{figure}

\paragraph{Preprocessing}
We integrate weekly U.S. \textit{Billboard Hot 100} chart data 
(1960s--2010s) with \textit{Spotify} audio features for the 
all-time top-10 artists. Rank 1--10 refers to positions on 
Billboard's weekly consumption-based chart, with eligibility 
governed by Billboard. Artists are ranked by the following score:
$\left( \sum_{\text{songs}} \frac{\text{weeks}_i}{\text{avg\_rank}_i} \right)
\times \log\!\bigl(1 + |\text{Songs}|\bigr)$,
where $\frac{\text{weeks}_i}{\text{avg\_rank}_i}$ represents each 
song's contribution and $|\text{Songs}|$ denotes the artist's 
distinct song count. We use five interpretable Spotify audio 
features: valence (musical positivity), energy (perceived intensity 
and activity), danceability (suitability for dancing), acousticness 
(likelihood of an acoustic recording), and liveness (likelihood of 
a live performance).

\subsection{Visual Encoding}

The system comprises four coordinated views (Figures~\ref{fig:main}--\ref{fig:detail}), illustrated using Michael Jackson's 1990s profile as a running example.

\paragraph{\circled{1} Artist--Decade Bubble Chart}
The main view supports (T1) by visualizing artist performance 
trajectories across six decades. Each bubble represents an 
artist--decade pair, with x-position encoding decade, y-position 
encoding artist rank, size encoding total chart appearances, and 
color intensity encoding the decade-specific performance score,
$\sum_{\text{songs}} \frac{\text{weeks}_i}{\text{avg\_rank}_i} 
\;\times\; |\text{Songs}|_d,$
where $|\text{Songs}|_d$ is the distinct song count within that 
decade. Connecting lines trace each artist's career path, 
allowing users to identify sustained activity or era-specific 
prominence. In the running example, Michael Jackson's 1990s bubble 
reflects 178 chart appearances across 11 distinct songs, 
a contraction from his peak 1980s activity 
(266 appearances) that the connecting line makes immediately apparent.

\vspace{-0.05cm}

\paragraph{\circled{2} Quadrant Trajectory Map}
Each artist--decade pair is plotted in a 2D space defined by two 
complementary metrics. The x-axis encodes \emph{shape 
similarity}---centered cosine similarity between the artist's mean 
audio feature vector and the era centroid---capturing directional 
correspondence. The y-axis encodes \emph{contrast ratio} 
($\sigma_\text{artist} / \sigma_\text{era}$), capturing whether the 
artist's profile is more extreme or uniform than the era norm. A 
Pearson correlation test over all 33 artist--decade pairs showed no 
significant linear association between the two axes ($r = -0.19$, 
$p = .30$), suggesting that they capture distinct dimensions. The 
space is partitioned by a median-based boundary on shape similarity 
and a theoretically grounded boundary on contrast ratio 
($\sigma_\text{artist}/\sigma_\text{era} = 1.0$), defining four 
quadrants: \textit{Amplified Conformist} (high shape, high 
contrast), \textit{Smoothed Conformist} (high shape, low contrast), 
\textit{Polarized Maverick} (low shape, high contrast), and 
\textit{Muted Maverick} (low shape, low contrast). Lines connect 
each artist's consecutive decade positions, enabling trajectory 
tracing (T1, T3). In the running example, Michael Jackson's 1990s 
position (shape\,=\,0.995, contrast\,=\,1.313) places him in the 
\textit{Amplified Conformist} quadrant, indicating strong 
directional alignment with the era baseline and amplified stylistic 
intensity---shifted upward from his 1980s position 
(shape\,=\,0.984, contrast\,=\,0.995), which sat near the quadrant 
boundary.

\vspace{-0.05cm}

\paragraph{\circled{3} Song Table and \circled{4} Audio Profiles}
Clicking a bubble in the main view reveals two coordinated panels.
The left panel displays a ranked table with average rank, peak rank, 
and weeks on chart, supporting (T2). In the running example, the 
table lists \textit{Remember The Time} (avg.\,rank\,19.0, 
peak\,3, 20\,weeks) and \textit{You Are Not Alone} 
(avg.\,rank\,22.5, peak\,1, 20\,weeks) among the top entries. 
The right panel presents three components. First, an \emph{Alignment 
Badge} supporting (T3) displays shape similarity and contrast ratio 
and classifies the artist--decade pair into one of the four 
quadrants defined above. Second, an \emph{Artist--Era Deviation} 
chart---a diverging bar chart supporting (T3)---displays how the 
artist's audio features in the selected decade deviate from the era 
average across five dimensions (valence, energy, danceability, 
acousticness, liveness). In the running example, the deviation chart 
reveals that despite near-perfect directional alignment, Michael 
Jackson's 1990s profile exhibits pronounced positive deviations in 
energy ($+0.11$) and danceability ($+0.11$)---the amplification 
pattern suggested by the quadrant position but not decomposed into 
individual features. Selecting a song row updates the right panel 
with a \emph{Song Signature}: a diverging bar chart showing per-song 
deviations from the era average, and a radar chart overlaying 
decade, artist, and song-level profiles for direct comparison, 
supporting (T2) and (T3).
\section{Expert Case Study and Discussion}

We conducted an open-ended, think-aloud session with a domain 
expert in commercial music production and chart dynamics (5+ years 
of experience). The participant noted that the quadrant-based 
framing surfaces distinctions that are difficult to capture using 
conventional chart metrics or genre labels alone, pointing to 
Madonna (1980s--2000s) as consistently occupying the 
\textit{Amplified Conformist} quadrant and Elton John as remaining 
within the \textit{Smoothed Conformist} quadrant.

\vspace{-0.01cm}

\paragraph{Limitations and Future Work}
This study relies on a single-participant exploratory session and a 
dataset limited to ten artists, constraining the generalizability 
of the findings. However, the core metrics---centered cosine 
similarity and contrast ratio---are independent of corpus size, 
feature set, or temporal granularity and can be applied to larger 
artist pools, alternative audio descriptors, or sub-decade 
intervals. Future work may evaluate scalability through 
multi-participant studies, expanded datasets, cross-artist analysis, 
and trajectory pattern mining to identify common career pathways.



\bibliographystyle{abbrv-doi}

\bibliography{template}
\end{document}